\documentclass[12pt]{iopart}

\usepackage{cite}
\usepackage{graphicx}
\usepackage{subcaption}
\usepackage{lineno,hyperref}
\usepackage{booktabs}
\expandafter\let\csname equation*\endcsname=\relax
\expandafter\let\csname endequation*\endcsname=\relax
\usepackage{amsmath,amssymb,amsfonts}
\usepackage{bm}
\usepackage{multirow}

\begin{document}

\title[PaMMA-Net: Plasmas magnetic measurement evolution model]
{PaMMA-Net: Plasmas magnetic measurement evolution based on data-driven incremental accumulative prediction}

\author{Yunfei Ling$^1$, Zijie Liu$^{2,3}$, Jun Du$^{1}$, Yao Huang$^3$, Yuehang Wang$^3$, Bingjia Xiao$^{3}$, Xin Fang$^4$}

\address{$^1$ University of Science and Technology of China, Hefei 230027, China}
\address{$^2$ Institute of Energy, Hefei Comprehensive National Science Center, Anhui, Hefei, 230031, China}
\address{$^3$ Institute of Plasma Physics, Hefei Institutes of Physical Science, Chinese Academy of Sciences, Hefei 230031, China}
\address{$^4$ iFLYTEK Research, Hefei 230026, China} 
\ead{\mailto{dujun@ustc.edu.cn}, \mailto{lingyunfei@mail.ustc.edu.cn}}
\vspace{10pt}

\begin{indented}
\item[]January 2025
\end{indented}

\begin{abstract}
An accurate evolution model is crucial for effective control and in-depth study of fusion plasmas. Evolution methods based on physical models often encounter challenges such as insufficient robustness or excessive computational costs. Given the proven strong fitting capabilities of deep learning methods across various fields, including plasma research, this paper introduces a deep learning-based magnetic measurement evolution method named PaMMA-Net (\textbf{P}l\textbf{a}sma \textbf{M}agnetic \textbf{M}easurements Incremental \textbf{A}ccumulative Prediction Network). This network is capable of evolving magnetic measurements in tokamak discharge experiments over extended periods or, in conjunction with equilibrium reconstruction algorithms, evolving macroscopic parameters such as plasma shape. Leveraging a incremental prediction approach and data augmentation techniques tailored for magnetic measurements, PaMMA-Net achieves superior evolution results compared to existing studies. The tests conducted on real experimental data from EAST validate the high generalization capability of the proposed method.
\end{abstract}

\vspace{2pc}

\maketitle

\section{Introduction}

Fusion energy presents three major advantages: abundant reserves, minimal environmental impact, and a high degree of safety, making it the ideal energy source for humanity. One of the most important ways to achieve controlled fusion is to generate and confine high-temperature plasma through tokamak devices~\cite{ongena2016magnetic}. The continuous development of numerical simulation methods for fusion plasmas in recent decades has demonstrated that establishing a model capable of accurately describing plasma behavior under the control of tokamak devices is crucial for a better understanding and control of plasma dynamics. The integration of numerical simulations, theoretical analyses, and experimental measurements facilitates the verification of the feasibility of tokamak experimental configurations, the development of advanced tokamak operational modes, and the effective control of plasma behavior.

The evolution methods for fusion plasmas could be categorized into two approaches: physics-driven methods and data-driven methods~\cite{batchelor2007simulation}. Physics-driven methods entail partial differential equations, sophisticated numerical methods, and simplifying assumptions. Data-driven methods, on the other hand, primarily consist of empirical equations and deep learning techniques. Physics-driven methods boast excellent interpretability, but for first-principles models, their accuracy hinges on the completeness of the physical processes involved, making it challenging to strike a balance between efficiency and precision. When efficiency is constrained, it becomes difficult to account for all physical phenomena comprehensively. In contrast, data-driven methods generally feature models and numerical methods of lower complexity, resulting in faster computation speeds. Furthermore, as data-driven methods are grounded in raw data, they could inherently accommodate a broader range of physical phenomena.

Deep neural networks, as a data-driven approach to modeling complex parametric relationships, have made significant progress in the last decade. Especially since the Transformer architecture~\cite{vaswani2017attention} has been proposed and extensively studied, deep neural networks have achieved impressive results in time series modeling and prediction. It has been widely used in many fields, such as natural language generation~\cite{achiam2023gpt}, sequence prediction in scenarios like power~\cite{sharadga2020time}, finance~\cite{tang2022survey}, etc. In the field of plasma control, deep neural networks have also been applied in many research directions. Such as disruption prediction~\cite{guo2023disruption}, equilibrium reconstruction~\cite{lao2022application}, discharge prediction~\cite{wan2022east}. It is generally believed that deep neural networks, takes advantage of their strong generalization, could produce better predictions than physical models. By self-supervised training on rich experimental data, deep neural networks can be generalized to some unseen experimental configurations.

Compared to univariate, time-independent regression tasks, the long-sequence plasma magnetic measurement evolution is a significantly more challenging problem. It demands high precision in modeling the plasma response and requires capturing long-range dependencies with strong generalization ability. Both the inputs and outputs involve multiple modalities, each with a wide dynamic range and substantial distributional differences. These characteristics present greater challenges in the design of magnetic measurement evolution models. In contrast to some existing data-driven methods for plasma discharge prediction, this paper introduces targeted improvements to the model based on the characteristics of magnetic measurement signals and endeavors to make better use of observed measurements. Consequently, it achieves long-sequence magnetic measurement evolution with fewer inputs.

To efficiently model plasma behavior, this paper adopts a two-phase plasma evolution approach combining long-sequence magnetic measurements evolution with equilibrium reconstruction. Our primary focus is on magnetic measurement prediction, as it embodies the most authentic state information of the plasma, encapsulating the influence of all factors such as electromagnetic control, heating and material injection, and plasma transport. Consequently, this method possesses inherent advantages in terms of accuracy and generalization. Furthermore, through equilibrium reconstruction, the magnetic measurement evolution yields a rich evolution of reconstructed signals. Leveraging a data-driven approach, our method models the autocorrelation of plasma measurement signals across different time points and their cross-correlation with signals such as coil currents. The supervised-trained model achieves high accuracy in plasmas magnetic measurements evolution. Therefore, this paper introduces a novel approach, termed PaMMA-Net (\textbf{P}l\textbf{a}sma \textbf{M}agnetic \textbf{M}easurements Incremental \textbf{A}ccumulative Prediction Network), which leverages deep neural networks to predict plasma magnetic measurements. By employing incremental accumulative prediction, appropriate model design, and physically consistent data augmentation techniques, the proposed method attains superior predictive performance compared to other neural networks.

The contributions of this paper can be summarized as follows: 1) A novel data-driven approach, named PaMMA-Net, is proposed for long-sequence plasma magnetic measurement evolution. Its core design, incremental accumulative prediction, has been demonstrated to enhance the performance of various neural networks in predicting magnetic measurements. 3) Experiments on real-recorded data verify the prediction accuracy of the proposed method surpasses known data-driven models with similar computational costs, especially under the scenario of long-term prediction.

\section{Related work}
\subsection{Physics-driven plasma modeling}
Physics-driven models are typically integrated versions of first-principles-based physical models~\cite{falchetto2014european}, which often employ separate modules to address sub-problems such as equilibrium reconstruction \cite{zwingmann2003equilibrium, zwingmann2008plasma}, refinement \cite{westerhof1992relax, lutjens1996chease}, and linear MHD stability calculations \cite{huysmans2001modeling, liu2000feedback}. However, the accuracy of these methods hinges on the completeness of the involved physical processes, posing a challenge in balancing efficiency and precision. Specifically, when prioritizing efficiency, it becomes difficult to encompass all physical phenomena. Conversely, excessive simplification of the model and overly stringent assumptions may lead to uncertainties in the results.

\subsection{Data-driven plasma modeling}
In recent years, data-driven methods have been extensively applied to various plasma modeling tasks, including disruption prediction \cite{hu2021real, guo2021disruption, yoshino2003neural}, electron temperature profile estimation~\cite{clayton2013electron}, discharge estimation \cite{wan2021experiment, wan2023machine}, last closed-flux surface evolution~\cite{wan2024predict}, equilibrium reconstruction \cite{lao2022application, joung2019deep}, control plasma \cite{yang2020design, wakatsuki2019safety}, reinforcement learning-informed magnetic field control \cite{seo2021feedforward, degrave2022magnetic} and instability avoidance~\cite{seo2024avoiding}.

As a field dedicated to studying the temporal relationships among variables, time series forecasting has already provided some effective modeling approaches for data-driven plasma modeling. LSTM~\cite{hochreiter1997long}, TCN~\cite{bai2018empirical} and DeepAR~\cite{salinas2020deepar}, Timesnet~\cite{wu2022timesnet} use CNNs and RNNs to predict future trends. MLP-based models such as N-HiTS~\cite{challu2023nhits}, N-BEATS~\cite{oreshkin2019n}, NLinear~\cite{zeng2023transformers}, and DLinear~\cite{zeng2023transformers} maintain high accuracy even with fewer parameters. Due to the breakthrough in computer vision and natural language processing, Transformer-based models have also seen widespread adoption in TSF. Informer~\cite{zhou2021informer}, FEDformer~\cite{zhou2022fedformer}, Autoformer~\cite{wu2021autoformer}, and PatchTST~\cite{nie2022time}, itransformer~\cite{liu2023itransformer} are able to effectively capture key information in the sequence to significantly improve accuracy.

It is noteworthy that the magnetic measurement evolution task defined in this paper can be categorized as a multi-horizon forecasting task~\cite{lim2021temporal} because some relevant future inputs are already known when predicting future sequences. Such tasks require modeling the relationship among the future sequence, the observed sequence and known future inputs simultaneously.

\section{Methodology}
\subsection{Magnetic measurements evolution}
\label{section:m2m}
Typically, Tokamak discharge experiments measure the magnetic field intensity and flux at specific locations using pickup coils and flux loops. Subsequently, the magnetic flux in all regions of the vacuum chamber is calculated using equilibrium reconstruction algorithms~\cite{lao1990equilibrium}. Finally, based on the magnetic field data, the coil currents are controlled in real-time to maintain the plasma state according to control objectives. In alignment with this standard procedure, the plasma magnetic measurement evolution model defined in this paper does not directly predict the magnetic flux in all regions. Instead, it simulates the measurement process of Tokamak discharge experiments: predicting future sensor measurements based on a segment of observed data and the coil currents throughout the experiment. This task design offers two key advantages: it avoids introducing additional errors from the equilibrium reconstruction process, and it more directly simulates the experimental measurement process, thereby facilitating integration with other Tokamak simulation algorithms.

\subsubsection{Task definition}
In the context of data-driven multivariate regression, the magnetic measurement evolution task begins with the collection of a dataset, which comprises $\mathrm{K}$ independent time series $\bm{U}^{(k)}\in\mathbb{R}^{\mathrm{C} \times \mathrm{T}_k}$. Each time series encapsulates a complete discharge experiment, characterized by $\mathrm{T}_k$ time-steps and $\mathrm{C}$ signal channels, denoted by $\bm{U}^{(k)}=\{\bm{u}^{(k)}_\mathrm{1},...,\bm{u}^{(k)}_{\mathrm{T}_k}\}$, $\bm{u}^{(k)}_{t}=\{u_1,...,u_\mathrm{C}\}\in\mathbb{R}^{\mathrm{C}}$. Notably, the number of signal channels remains constant across all series, whereas the number of time-steps may vary. The signal channels are categorized into two groups: the first $\mathrm{D}$ channels, referred to as magnetic measurement or state signals $\bm{y}_t=\{u_1,...,u_\mathrm{D}\}\in\mathbb{R}^{\mathrm{D}}$, and the remaining $\mathrm{E}$ channels, designated as control signals $\bm{z}_t=\{u_\mathrm{D},...,u_\mathrm{C}\}\in\mathbb{R}^{\mathrm{E}}$. For any given time $t$, the signals preceding it are designated as the observed signals $\bm{X}_{t-\mathrm{M}-1:t}=\{\bm{u}_{t-\mathrm{M}-1},...,\bm{u}_{t}\}\in\mathbb{R}^{\mathrm{C} \times \mathrm{M}}$, the subsequent magnetic measurement signals constitute the evolution target $\bm{Y}_{t+1:t+\mathrm{N}}=\{\bm{y}_{t+1},...,\bm{y}_{t+\mathrm{N}}\}\in\mathbb{R}^{\mathrm{D} \times \mathrm{N}}$, and the subsequent control signals are termed known future inputs $\bm{Z}_{t+1:t+\mathrm{N}}=\{\bm{z}_{t+1},...,\bm{z}_{t+\mathrm{N}}\}\in\mathbb{R}^{\mathrm{E} \times \mathrm{N}}$, where $\mathrm{M}$ is the length of the observed input, $\mathrm{N}$ is the length of the  evolution result. In subsequent chapters, $\mathrm{M}$ and $\mathrm{N}$ are assigned values of 1000. The objective of the magnetic measurement evolution task is to predict the evolution target based on the observed input and known future input. Hence, the task could be defined as: 

\vspace{-2mm}
\begin{equation}
\bm{\widehat{Y}}_{t+1:t+\mathrm{N}}=f(\bm{X}_{t-\mathrm{M}-1:t},\bm{Z}_{t+1:t+\mathrm{N}})
\label{eq:task-def}
\end{equation}

where $\bm{\widehat{Y}}_{t+1:t+\mathrm{N}}$ is the predicted output of the future $\mathrm{N}$ steps, $f(\cdot)$ represents the proposed PaMMA-Net.

\subsubsection{Input and output variables}
Detailed input and output signals used by our model and their number of channels are shown in Table \ref{tab:io}. Where the subscript $_{:t}$ of $BP_{:t}$ and $FL_{:t}$ indicates that the prediction starts at time $t$, and the observed signals referring to signals prior to time $t$. The subscript $_{t:}$ denotes signals that are predictions for future measurements. As for the known future inputs without a time subscript, they are known at any given moment. The model utilizes these signals while ensuring causality is maintained.

\begin{table}[tbp] 
    \centering
    \vspace{1mm}
    \caption{Physical meaning and number of channels of chosen input and output signals for magnetic measurement evolution task. Including observed signals, known future input and evolution target} 
    \vspace{1mm}
    \label{tab:io} 
    \renewcommand{\arraystretch}{1.0}
    \setlength\tabcolsep{14.0pt}  
    \resizebox{10cm}{!}{
        \begin{tabular}{c|c|c} 
            \toprule 
            {Name} & {Physical meaning} & {\# Channels} \\
            \cline{1-3}
            \multicolumn{3}{c}{Observed signals}\\
            \cline{1-3}
            {$BP_{:t}$} & {Magnetic surface signal} & {37} \\
            {$FL_{:t}$} & {Magnetic flux signal} & {35} \\
            \cline{1-3}
            \multicolumn{3}{c}{Known future input}\\
            \cline{1-3}
            {$I_\mathrm{p}$} & {Plasma current} & {1} \\
            {$V_\mathrm{loop}$} & {Loop voltage} & {1} \\
            {$W_\mathrm{MHD}$} & {Plasma stored energy} & {1} \\
            {$\beta_\mathrm{p}$} & {Poloidal beta} & {1} \\
            {$l_\mathrm{i}$} & {Internal inductance} & {1} \\
            {$PF_\mathrm{cmd}$} & {Poloidal field coils voltage} & {12} \\
            {$IC$} & {In-vessel coil No.1 voltage} & {1} \\
            \cline{1-3}
            \multicolumn{3}{c}{Evolution target}\\
            \cline{1-3}
            {$BP_{t:}$} & {Magnetic surface signal} & {37} \\
            {$FL_{t:}$} & {Magnetic flux signal} & {35} \\
            \bottomrule
        \end{tabular}
    }
\vspace{-3mm}
\end{table}

In accordance with the task definition outlined in the previous section, this paper adopts a total of C = 90 signal channels. The magnetic measurements consist of 37-dimensional $BP$ signals and 35-dimensional $FL$ signals, amounting to D = 72 signal channels in total. The known future inputs encompass PCS commands and several macroscopic signals, totaling E = 18 signal channels. This selection is made with the intention of utilizing the minimal number of signals necessary to achieve plasma evolution, thereby facilitating the construction of a highly generalized evolution model for magnetic measurements.

\subsection{PaMMA-Net}
\label{section:iat}
\begin{figure*}[t!]
    \centering
    \includegraphics[width=15.5cm]{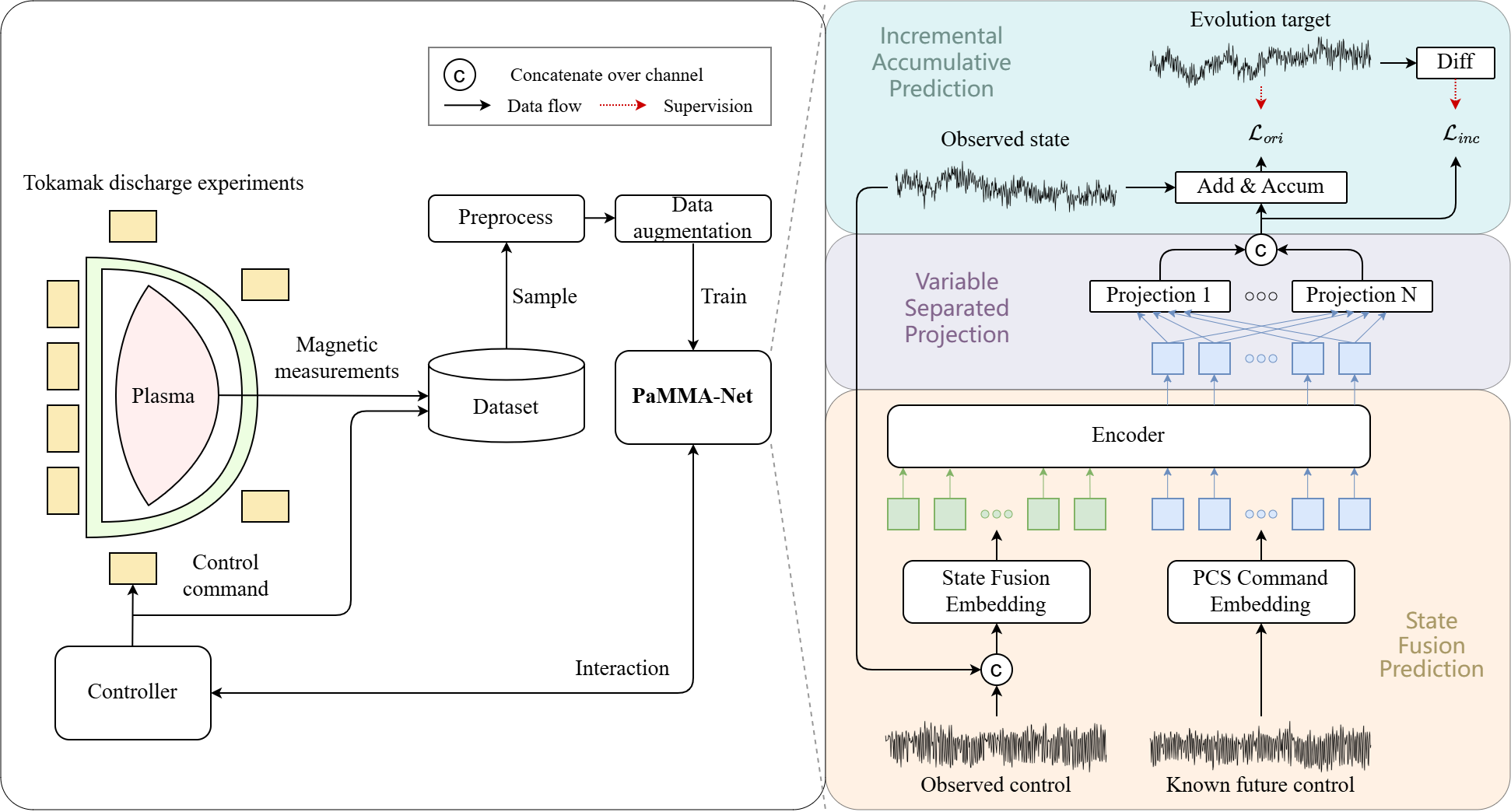}
    \caption{Workflow and model architecture of PaMMA-Net. The left half presents the process of collecting magnetic measurements and control commands from tokamak discharge experiments to form a dataset, as well as the subsequent steps of preprocessing and data augmentation applied to this dataset for training purposes.. The right half displays detailed architecture of PaMMA-Net, including state fusion prediction, variable separation projection, and incremental accumulative prediction.}
    \vspace{-2mm}
    \label{fig:Archi}
\end{figure*}

This paper introduces PaMMA-Net, which imposes additional supervision on the increments of evolution targets and efficiently models the relationship between magnetic measurements and control signals using canonical components. Consequently, it achieves high-performance magnetic measurements evolution in the field of tokamak discharge prediction. The overall model architecture and data processing methodology of PaMMA-Net are illustrated in Fig. \ref{fig:Archi}.

The left half of Fig. \ref{fig:Archi} provides a concise overview of the data acquisition and processing pipeline for the proposed method. Initially, the required signals are collected from a vast array of tokamak discharge experiments to form a comprehensive dataset. Subsequently, signal segments are sampled from this dataset and, after undergoing preprocessing and data augmentation, are utilized for the training of PaMMA-Net. Given the significant variation in dynamic ranges among signals from different channels and modalities, the primary preprocessing step involves applying 01 standardization on each signal channel. The resultant well-converged magnetic measurement evolution model could then be employed for offline interaction with the controller.

The right half of Fig. \ref{fig:Archi} displays the primary structure of PaMMA-Net. As described in the task definition, the model evolves magnetic measurements based on observed signals and known future inputs. From input to output, it comprises three components: State fusion prediction, Variable separation projection, and Incremental accumulative prediction. Subsequent sections will delve into each of these components individually.

\subsubsection{Incremental accumulative prediction}
The term "increment" refers to the temporal difference of the prediction target, denoted $\Delta\bm{y}_{t}$. Our model first predicts the increment and then, through a non-parametric accumulation layer, converts the increment back to the prediction target. Both the increment and the predicted measurements serve as targets for supervised learning. Based on this design, the Eq. \ref{eq:task-def} can be reformulated as follows.

\vspace{-2mm}
\begin{equation}
\begin{split}
&\Delta\bm{Y}_{t+1:t+\mathrm{N}}=\{\Delta\bm{y}_{t+1},...,\Delta\bm{y}_{t+\mathrm{N}}\}, \Delta\bm{y}_{t}=\bm{y}_{t}-\bm{y}_{t-1} \\
&\Delta\bm{\widehat{Y}}_{t+1:t+\mathrm{N}}=f(\bm{X}_{t-\mathrm{M}-1:t},\bm{Z}_{t+1:t+\mathrm{N}}) \\
&\bm{\widehat{y}}_{t+n}=\bm{y}_{t} + \sum^{n}_{\tau=1} \Delta\bm{\widehat{y}}_{t+\tau}
\end{split}
\label{eq:diff}
\end{equation}

\begin{figure}[htbp]
  \centering
  \subfloat[\label{fig:4a}]{
    \includegraphics[width=7.7cm]{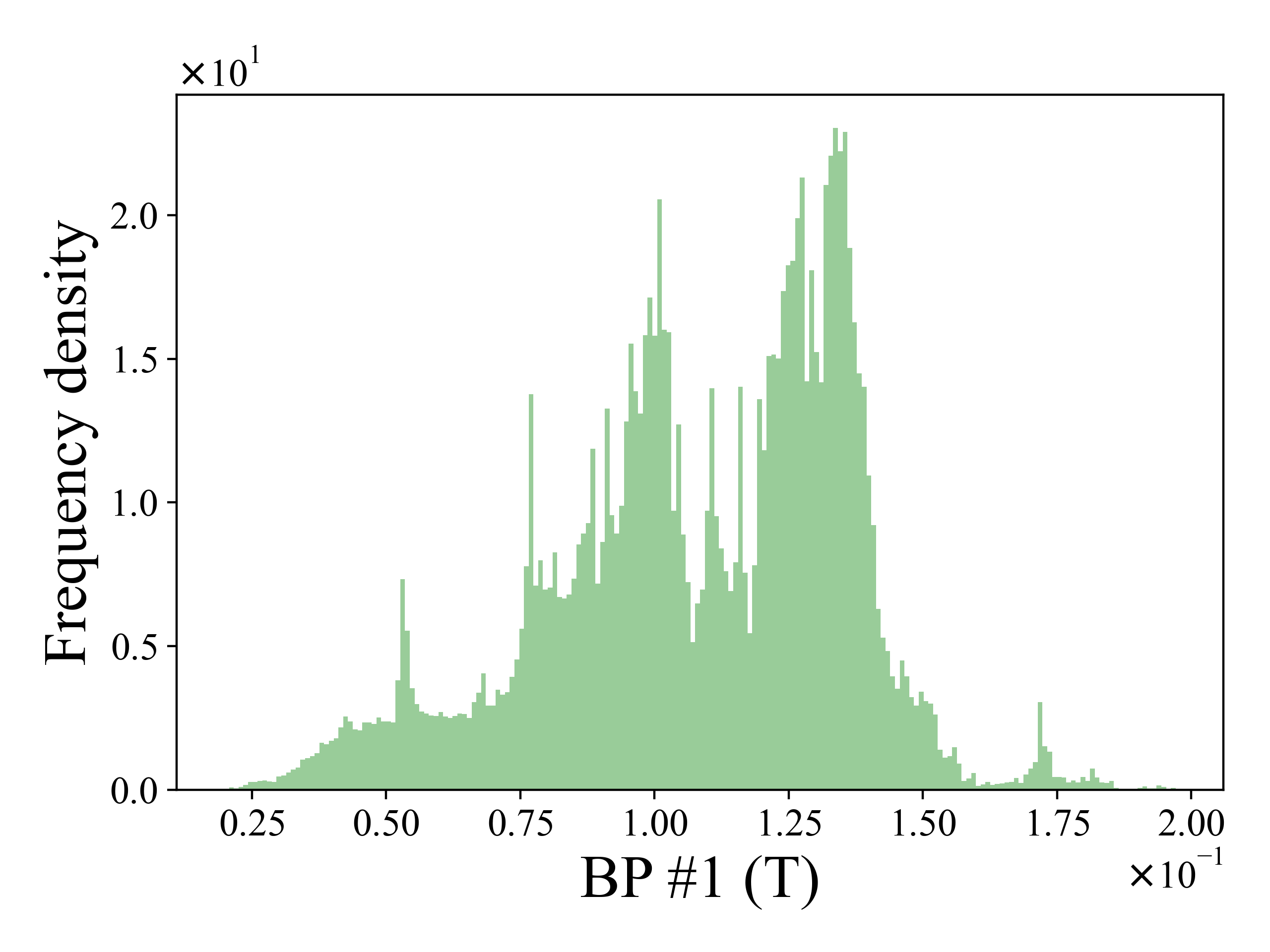}
    \vspace{-1mm}
  }
  \subfloat[\label{fig:4b}]{
    \includegraphics[width=7.7cm]{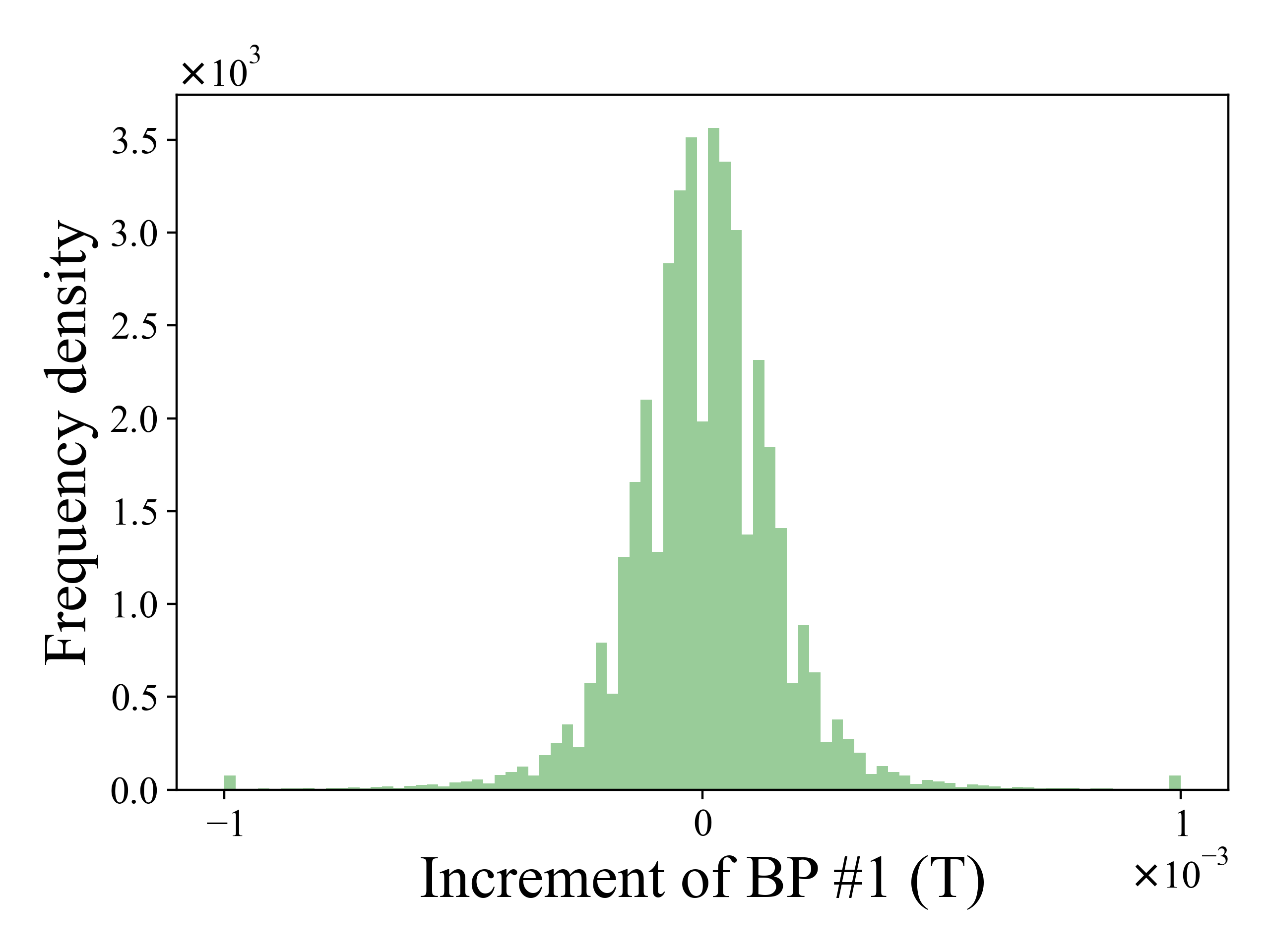}
    \vspace{-1mm}
  }
  \vspace{-1mm}
  \caption{(a) Distribution of $BP$ channel 1 within the dataset, (b) Distribution of the increment of $BP$ channel 1 within the dataset.}
  \vspace{-4mm}
  \label{fig:diff_distr}
\end{figure}

The design of increment accumulative prediction stems from the following intuition: the non-normal distribution of the magnetic measurements $\bm{Y}$ versus the normal distribution of the incremental signals $\Delta\bm{Y}$, as illustrated in Fig. \ref{fig:diff_distr}. It is evident that the distribution of $\Delta\bm{y}_{i,t}$ aligns well with the normal distribution, whereas the distribution of $\bm{y}_{i,t}$ is the accumulative result of $\Delta\bm{y}_{i,t}$'s distributions, and the distribution of $\bm{Y}_i$ is the summation of $\bm{y}_{i,t}$'s distributions. Through such an accumulative process, the distribution of $\bm{Y}_i$ exhibits complex non-normality. We believe that associating the evolution task with the prediction of normally distributed variables could facilitate better convergence of the model. Furthermore, the dynamic range of $\Delta\bm{Y}$ being much smaller than $\bm{Y}$ is conducive to promoting the model's ability to capture finer-grained plasma behaviors, thereby enhancing prediction accuracy.

\subsubsection{State fusion prediction}
As defined in Sec. \ref{section:m2m}, the proposed evolution model incorporates observed magnetic measurements as input, thereby implicitly modeling the relationships among various signal channels and furnishing an initialization for incremental prediction. However, this state fusion design results in a difference in the number and physical meanings of input channels before and after time t. The crucial challenge lies in unifying these into a format that can be processed by the attention module. The method adopted in this paper could be defined by the following equations:

\vspace{-2mm}
\begin{equation}
\begin{split}
\bm{h}^{\mathrm{S}}_{t} &= \mathrm{Concat}(\mathrm{MLP^{S}}(\bm{X}_{t}), \bm{X}_{t})+\mathrm{PE}_t  \\
\bm{h}^{\mathrm{C}}_{t} &= \mathrm{Concat}(\mathrm{MLP^{C}}(\bm{X}_{t}), \bm{X}_{t})+\mathrm{PE}_t  \\
\bm{H}^{0} &= \mathrm{Concat}(\bm{H}^{\mathrm{S}}, \bm{H}^{\mathrm{C}})
\end{split}
\label{eq:embed}
\end{equation}

where $\bm{H}^{\mathrm{S}}=\{\bm{h}^{\mathrm{S}}_{t-\mathrm{M}-1}, ..., \bm{h}^{\mathrm{S}}_{t+\mathrm{N}}\}\in\mathbb{R}^{\mathrm{d_A} \times \mathrm{M+N}}$, $\bm{H}^{\mathrm{C}}=\{\bm{h}^{\mathrm{C}}_{t-\mathrm{M}-1}, ..., \bm{h}^{\mathrm{C}}_{t+\mathrm{N}}\}\in\mathbb{R}^{\mathrm{d_A} \times \mathrm{M+N}}$, $\mathrm{d_A}$ is the projected dimension. The superscript $0$ of $\bm{H}^{0}$ denotes the input of the first transformer layer. The superscript $\mathrm{S}$ and $\mathrm{C}$ are used to distinguish state fusion embedding and PCS command embedding. PE is the result of position embedding.

As described by Eq. \ref{eq:embed}, two embedding layers with the same basic components but independent parameters are designed, named state fusion embedding and PCS command embedding, to process the heterogeneous observed signals and known inputs. Both embedding layers incorporate token embedding and position embedding. Token embedding employs an MLP layer to tokenize the preprocessed and data-augmented inputs into tokens required for subsequent decoder layers. Specifically, to preserve the original features, the input is concatenated with the output of the MLP layer. An MLP is used here because, in this task, the data across different channels are strongly coupled. Due to its fully connected nature across all feature dimensions with relatively few parameters, an MLP could better explore inter-channel relationships. In contrast, convolutional-based embeddings have a smaller receptive field. Position embedding follows the design in~\cite{vaswani2017attention}. The results of token embedding and position embedding are summed to obtain the final embedding output.

In this paper, cascaded decoder layers are utilized as the core component for plasma modeling after the embedding layer. Each decoder layer follows a common architecture, consisting of a masked multi-head attention (MHA) and a feed-forward network (FFN) with residual connections. The cascaded decoder layers and the attention mechanism adopted in this paper are defined as follows:

\vspace{-2mm}
\begin{equation}
\begin{split}
\bm{H}^{l+1} &= \mathrm{DecoderLayer}(\bm{H}^{l})  \\
\mathrm{A}(\bm{Q},\bm{K},\bm{V}) &= (\mathrm{Softmax}(\frac{\bm{Q}\bm{K}^{\top}}{\sqrt{\mathrm{d_A}}})+\bm{\mathrm{M}})\bm{V}
\end{split}
\label{eq:decoder}
\end{equation}

where queries, keys and values $\bm{Q}$, $\bm{K}$, $\bm{V}\in\mathbb{R}^{\mathrm{d_A} \times \mathrm{M+N}}$ are obtained by adopting linear projections on input $\bm{H}^{l}$, $\bm{\mathrm{M}}$ is causal mask.

The attention module employed in this paper utilizes causal masking rather than full attention to prevent the simultaneous acquisition of multi-step control signals, thereby ensuring the model possesses the capability for single-step autoregressive inference. The adoption of full attention would lead to leakage of contextual information, resulting in a mismatch between training and inference phases.

\subsubsection{Variable separated projection}
After the decoder layers, a special mapping layer is placed, referred to the variable separated projection. It is defined as follows:

\vspace{-2mm}
\begin{equation}
\begin{split}
\bm{\widehat{y}}_{t}=\mathrm{Concat}(\mathrm{MLP}^{\mathrm{BP}}(\bm{h}^{\mathrm{L}}_{t}), \mathrm{MLP}^{\mathrm{FL}}(\bm{h}^{\mathrm{L}}_{t}))
\end{split}
\end{equation}

The projector takes into account the different modalities of $BP$ and $FL$, applying parameter separated projections to them. Two distinct MLP layers form the variable separation projector. Both of their inputs stem from the outputs of the decoder layers, but their outputs separately model the relationships between $BP$, $FL$, and the latent variables. Compared to directly using a wider fully connected layer, the variable separated projection utilizes fewer parameters and achieves faster convergence. An alternative, more radical design is the channel separated projection, which adopts parameter separated projections for each signal channel. However, in the context of our task, since the channels within $BP$ and $FL$ are highly coupled, the channel separated projection does not provide significant performance gains compared to the substantial computational overhead it incurs.

\subsection{Physically consistent data augmentation}
Considering the high cost of the Tokmak discharge experiment, this paper delves into a data augmentation approach grounded in simple rules, aiming to enrich signal diversity while serving as a regularization technique to mitigate the risk of model overfitting.

\begin{figure}[htbp]
    \centering

    \begin{subfigure}[b]{15.5cm}
        \centering
        \includegraphics[width=15cm]{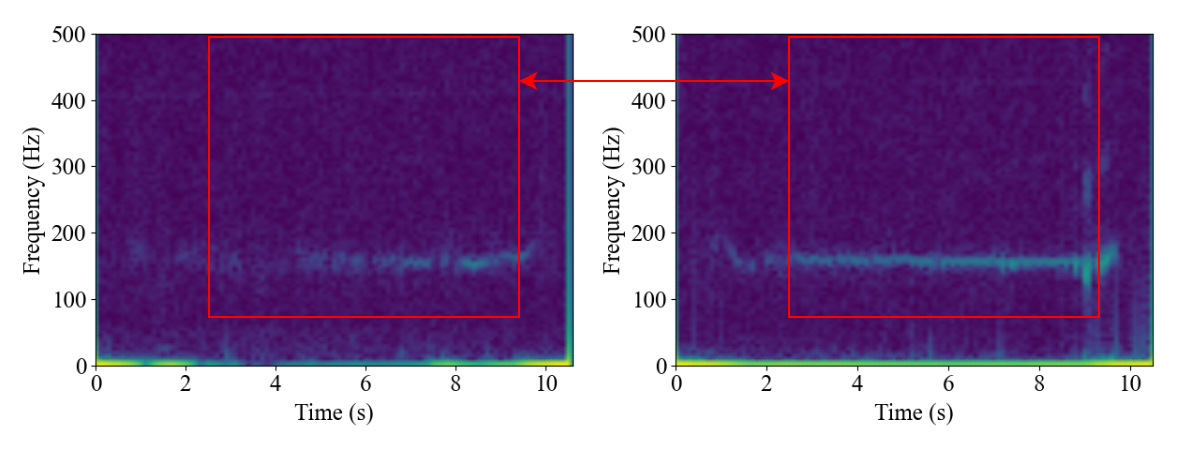}
        \caption{Spectrograms of $BP$ channel 1 in two shots.}
        \label{fig:Spectrum}
    \end{subfigure}
    
    \begin{subfigure}[b]{15.5cm}
        \centering
        \includegraphics[width=15cm]{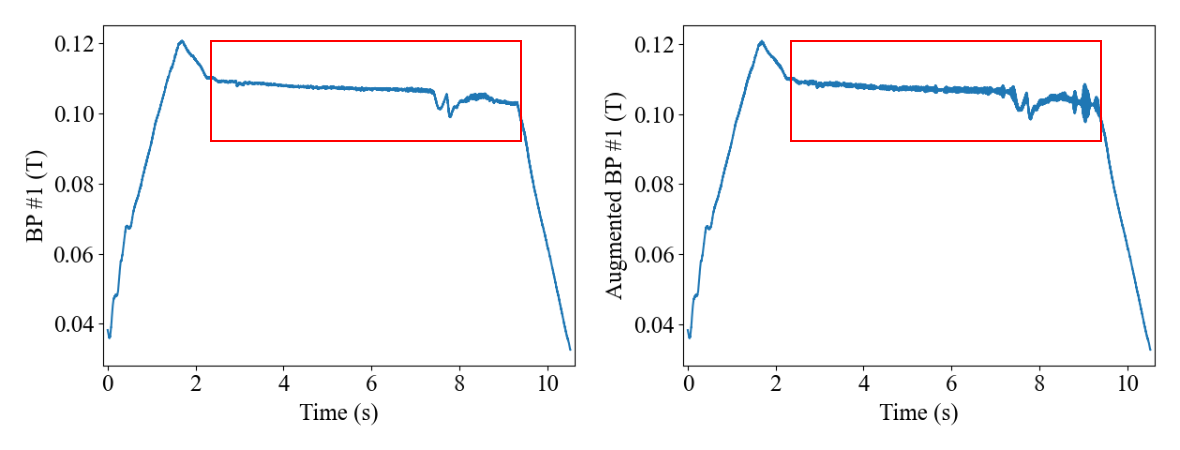}
        \caption{$BP$ channel 1 before and after augmentation}
        \label{fig:Spectrum-sub2}
    \end{subfigure}
    
    \caption{The principle and efficacy of the physically consistent data augmentation.}
    \vspace{-3mm}
    \label{fig:fig2}
\end{figure}

The spectrums of $BP$ and $FL$ are studied, as shown in Fig. \ref{fig:Spectrum}. It could be found that the signal is mostly in the fundamental frequency part. In addition, there are power supply harmonics around 160Hz, and the generation of this frequency component is relatively random. Therefore, CutMix~\cite{yun2019cutmix} is performed on the high-frequency part of the Short-Time Fourier Transform spectrums of the magnetic measurement signals to improve the richness of the data.

The aim of this data augmentation method is to generate a new time series $\widetilde{x}$ by combining the spectrums of two training samples $x_{A}$ and $x_{B}$. The generated training series $\widetilde{x}$ is used to train the model with its original loss function. The augmentation operation could be defined as:

\vspace{-2mm}
\begin{equation}
\widetilde{x}=\mathrm{istft}(\bm{\mathrm{M}}(\lambda) \odot \mathrm{stft}(x_{\mathrm{A}})+(\bm{1}-\bm{\mathrm{M}}(\lambda)) \odot \mathrm{stft}(x_{\mathrm{B}}))
\end{equation}

where $\widetilde{x}$, $x_{\mathrm{A}}$, $x_{\mathrm{B}}$ are single channel singles, $\mathrm{stft}(\cdot)$ and $\mathrm{istft}(\cdot)$ represent short-time Fourier transform and its inverted transformation. $\bm{\mathrm{M}}(\lambda)$ is a binary mask indicating the place to drop out and fill in from two spectrums, it has the same size as the spectrums, and a bounding box $\bm{B}(\lambda)$ inside of it. $\bm{B}(\lambda)=(r_x, r_y, r_h, r_w)$ indicating the cropping regions on the two spectrums. The region $\bm{B}$ in $\mathrm{stft}(x_{\mathrm{A}})$ is removed and filled in with the patch cropped from $\bm{B}$ of $\mathrm{stft}(x_{\mathrm{B}})$. $\bm{1}$ is a binary mask filled with ones. $\odot$ denotes element-wise multiplication. Following Mixup~\cite{zhang2017mixup}, the ratio $\lambda$ is sampled from the beta distribution $\mathrm{Beta}(\alpha,\alpha)$, for $\alpha \in (0,\infty)$.

The bounding box coordinates are determined according to:

\vspace{-2mm}
\begin{equation}
\begin{split}
& r_w=(\mathrm{W}-\mathrm{W}_0)(1-\lambda), \: r_h=\mathrm{H}-\mathrm{H}_0  \\
& r_x \sim \mathrm{Unif}(\frac{\mathrm{W}_0}{2},\mathrm{W}-\frac{\mathrm{W}_0}{2}-r_w), \: r_y=0
\end{split}
\end{equation}

where W, H are width and height of the spectrums, $\mathrm{H}_0$ represents the upper frequency limit that is not involved in the combination, which is 80 Hz here. $\mathrm{W}_0$ is the time range that does not participate in the combination, which is set to 0.1 W nearby the edge. $\mathrm{W}_0$ is set because of the edge effect of spectrum. $\mathrm{Unif}(\cdot)$ is uniform distribution.

Fig. \ref{fig:Spectrum-sub2} illustrates the enhancement effect on the original measurement signal through the application of the aforementioned transformation. The most significant changes are highlighted in red boxes. It is evident that this data augmentation method alters the high-frequency characteristics of the signal without altering its overall trend, thereby achieving an increase in data diversity. Both the original measurement signal and the augmented signal will be utilized for training purposes.It worth noticing that because this task is self-supervised training, the target and input are augmented simultaneously. 

\section{Experiment}
\subsection{EAST magnetic measurements dataset}
By screening the discharges with longer durations and higher sampling frequencies no from EAST experiments conducted between \#117823 and \#141775, two datasets were compiled in this study. The larger dataset, comprising 7671 shots for training, 199 shots for validation, and 830 shots for testing, was utilized to ascertain the model's evolutionary capability. The smaller one, consisting of 1000 shots for training, 50 shots for validation, and 100 shots for testing, was designated for an ablation study. Notably, the larger dataset encompasses the smaller one. The sampling rates per shot ranged from 500 to 10000 Hz, while the sampling durations varied between 5 and 50 seconds.

Furthermore, we have conducted a more granular segmentation of the dataset, dividing it into four mutually exclusive subsets based on two criteria: whether it is disruptive and whether it is in H-mode.  When allocating the training set, validation set, and test set, we ensured that the proportions of these four types of data remained largely consistent, thereby mitigating the risk of overfitting.

Before entering the model, we preprocessed the data as follows: In order to deal with inconsistent sampling frequency, we use linear interpolation method for up-sampling or down-sampling, and the sampling frequency of the signal is unified at 1000 Hz. Linear interpolation also fills in some values that are missing during measurement. Since our model could predict up to 1000 steps at a time, we use random clipping on the training set to obtain a fixed-length sequence, and intensive clipping with a window length of 2000 and stride of 1 on the test set.

\subsection{Experimental configuration}
\label{section:metrics}
\begin{figure}[htbp]
    \centering
    \includegraphics[width=11cm]{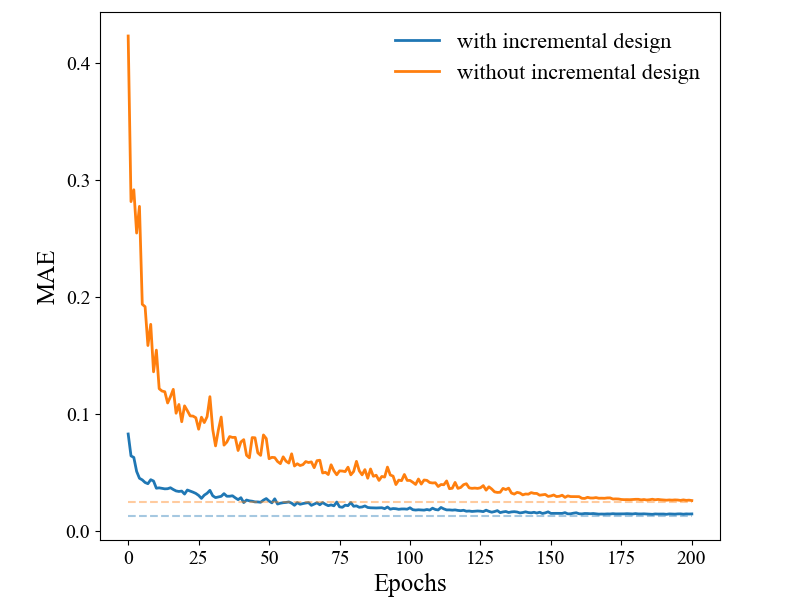}
    \caption{Learning behaviors of PaMMA-Net with and without incremental design.}
    \vspace{-3mm}
    \label{fig:learning curve}
\end{figure}

After obtaining the preprocessed and augmented data, we proceed with the training using the model and training configurations as outlined in Table \ref{tab:cfg}. The learning behavior of PaMMA-Net is depicted in Fig. \ref{fig:learning curve}.

\begin{table}[htbp]
    \centering
    \vspace{1mm}
    \caption{Utilized training configuration and model configuration} 
    \vspace{1mm}
    \label{tab:cfg} 
    \renewcommand{\arraystretch}{1.1}
    \setlength\tabcolsep{20.0pt}
    \resizebox{15cm}{!}{
        \begin{tabular}{cc|cc}
            \toprule 
            \multicolumn{2}{c}{Training configuration} & \multicolumn{2}{c}{Model configuration} \\
            \midrule
            {Learning rate} & {$1\times10^{-4}$} & {\# params} & {5.69M} \\
            {Optimizer} & {Adam} & {Input channels} & {90} \\
            {Scheduler} & {CosineAnnealing} & {Embedded dims} & {256} \\
            {\# epochs} & {200} & {\# decoder layers} & {6} \\
            \bottomrule
        \end{tabular}
    }
\vspace{-4mm}
\end{table}

We have delved into the impact of incremental accumulative prediction on the training of deep networks, conducting a comparative analysis with a baseline. Specifically, we contrast the validation MAE obtained during the training process with and without the incremental design. The results of this comparison are illustrated in Fig. \ref{fig:learning curve}.

It is evident that the incremental design facilitates a rapid convergence in the initial stages of model training. Upon completion of the first epoch, our model achieves a validation error significantly lower than the baseline. During the middle phase of training, our model exhibits smaller fluctuations in validation error. Ultimately, as indicated by the dashed line, our model converges to a lower level of validation error.

In this paper, the commonly used evaluation metric mean absolute error (MAE) is adopted. In addition, the metrics similarity and relative error are customized, which are denoted as Sim and Rela. Relative error measures the prediction accuracy in the form of percentage. Based on relative error, sim considers the fact that the mean and variance of each signal channel are different. Channels with high mean and low variance are easy to obtain low relative error. sim solves the negative impact of data distribution on evaluation results. Finally, the correlation index Corr is used to measure the trend similarity.The metrics for each time series are calculated as follows:

\vspace{-2mm}
\begin{equation}
\begin{split}
\mathrm{Rela} &= \frac{1}{\mathrm{T_{k}I}}\sum_{t}\sum_{i}
(1-\frac{|y_{i,t}-\widehat{y}_{i,t}|}{|y_{i,t}|}) \\
\mathrm{Sim} &= \frac{1}{\mathrm{T_{k}I}}\sum_{t}\sum_{i}
\frac{|y_{i,t}-\Bar{y}_{i}|}{|y_{i,t}-\Bar{y}_{i}|+|y_{i,t}-\widehat{y}_{i,t}|} \\
\mathrm{Corr} &= \frac{1}{\mathrm{I}}\sum_{i}
\frac{\sum_{t}(y_{i,t}-\Bar{y}_{i})(\widehat{y}_{i,t}-\Bar{\widehat{y}}_{i})}
{\sqrt{\sum_{t}(y_{i,t}-\Bar{y}_{i})^2(\widehat{y}_{i,t}-\Bar{\widehat{y}}_{i})^2}}
\end{split}
\end{equation}

It is noteworthy that we have adopted two testing methodologies in the subsequent text: non-autoregressive testing and auto-regressive testing. The distinction lies in their evolutionary approaches after time t; specifically, the non-autoregressive test evolves based on newly acquired magnetic measurements, whereas the auto-regressive test evolves according to the predictive outcomes. In essence, the non-autoregressive test assesses the model's evolutionary capability within 1000 steps, whereas the auto-regressive test evaluates the model's capability over a complete discharge process, typically exceeding 10,000 steps. Subsequent quantitative analyses are grounded on the non-autoregressive test, with separate case studies provided for the auto-regressive test.

\subsection{Forecasting result}
In this section, we first demonstrate the model's measurement-to-measurement prediction capability through a detailed statistical analysis of the prediction results for all shots in the test set. Subsequently, we illustrate the performance enhancements attributed to our model design by comparing it with several advanced deep learning models.

\begin{figure}[t!]
    \centering
    \includegraphics[width=12cm]{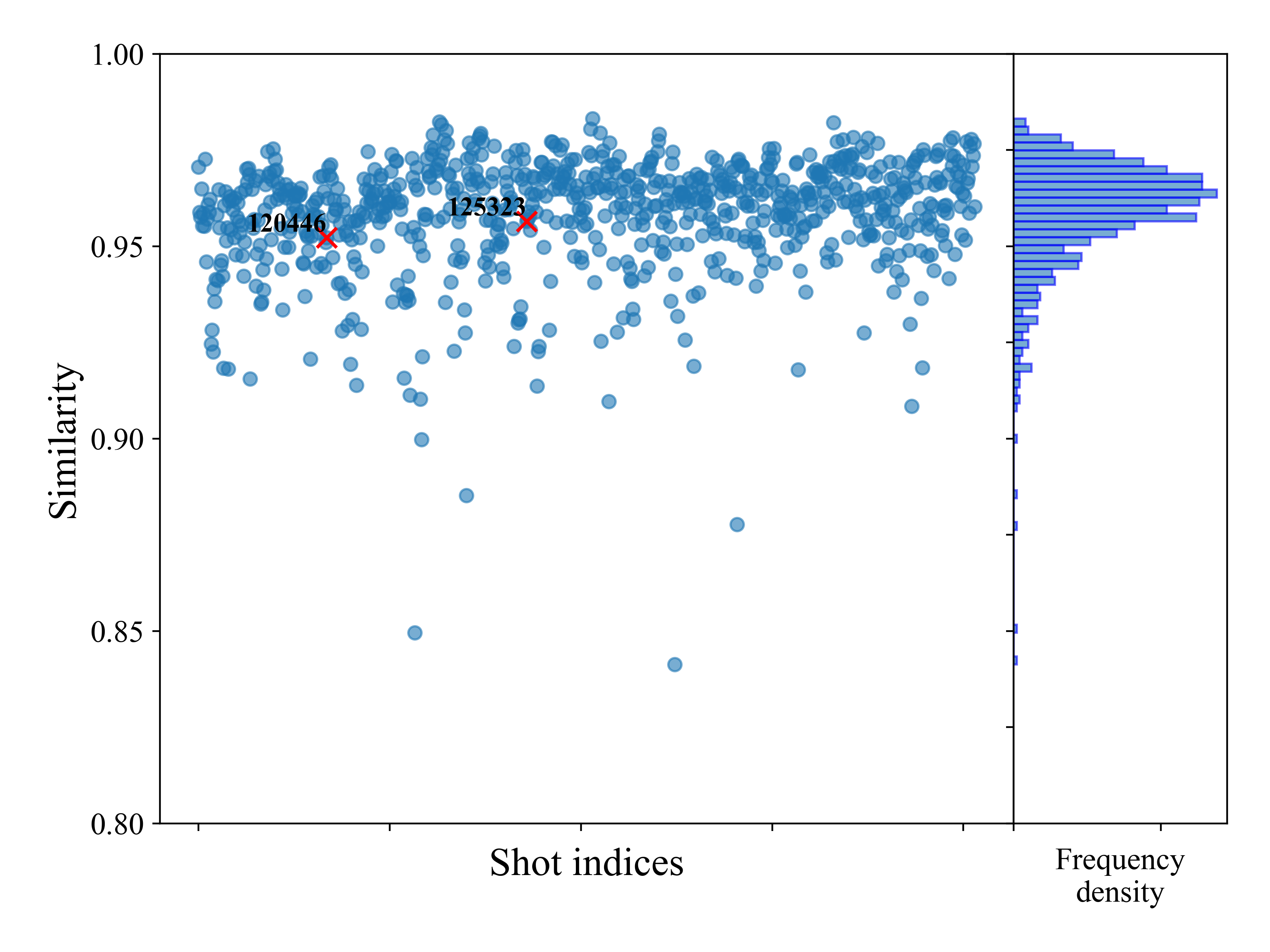}
    \vspace{-3mm}
    \caption{The evolution similarity metrics and its distribution of all discharges within the test set are presented. The left half depicts each discharge as an individual data point, while the right half illustrates the distribution using a horizontal histogram. Two discharges exhibiting moderate levels of evolution similarity have been selected and will be analyzed subsequently.}
    \vspace{-3mm}
    \label{fig:sim_distri}
\end{figure}

Fig. \ref{fig:sim_distri} presents a statistical analysis of the prediction similarity for all shots in the test set. The main scatter plot on the left visualizes the similarity for each shot, with each point representing an individual shot. The histogram on the right displays the distribution of all points from the left plot, where the vertical axis represents the similarity values, aligned with the scale of the main plot, and the horizontal axis represents the probability density. This figure reveals that the prediction similarity for the vast majority of shots falls between 90\% and 98\%, with an average similarity exceeding 95\%. Notably, even the worst predictions have a similarity around 85\%, which demonstrates the model's robust generalization ability to achieve good prediction results across shots with different experimental configurations. It is worth mentioning that, as described in Sec. \ref{section:metrics}, the similarity metric excludes the influence of signal mean and is, in most cases, a stricter metric than relative error. While achieving over 95\% similarity, the model also attains a Rela of over 99\%. The precise values of various metrics are reported in subsequent comparative experiments.

Subsequently, comparative experiments were conducted to evaluate the forecasting performance of our proposed model together with advanced time series forecasting methods. Several well-acknowledged methods are chose as our benchmark, including Transformer~\cite{vaswani2017attention}, Flowformer~\cite{wu2022flowformer}, LSTM~\cite{hochreiter1997long}, Non-stationary transformers~\cite{liu2022non}. Some models using temporal fusion design, like PatchTST~\cite{nie2022time} and iTransformer~\cite{liu2023itransformer}, are not validated because their non-causal nature is not suitable for our task.

In order to thoroughly validate the effectiveness of the proposed incremental accumulative prediction, two sets of experiments were conducted for each model involved in the comparison, with the incremental design being the sole variable. The experiments without the incremental design are denoted as “Original”, where the models utilized their basic structures as much as possible. Conversely, the experiments incorporating the incremental design are labeled as “+ IA”, where the models were augmented with a accumulative layer and additional supervision for predicting increments.

Comprehensive forecasting results are listed in Table \ref{tab:compare} with the best in bold. The lower MAE or higher Sim, Rela and Corr indicates the more accurate prediction result. It is evident that the proposed incremental accumulative prediction has brought performance enhancements to all model architectures in the magnetic measurement evolution task, albeit to varying degrees. As discussed in Sec. \ref{section:iat}, we attribute this improvement to a better representation of the slowly varying characteristics of magnetic measurements. Furthermore, due to the design of state fusion prediction and variable separated projection, our model achieves the best evolution results with fewer parameters. Specifically, it attains a Sim of 94.87\% and a Rela of 99.09\%.

\begin{table*}[t] 
    \centering
    \vspace{1mm}
    \caption{Evolution results of various models on the EAST dataset and performance promotion obtained by incremental accumulative prediction. For each model, two sets of test results are reported, depending on whether the incremental design is employed (denoted as “+ IA”) or not (denoted as “Original”).} 
    \vspace{1mm}
    \label{tab:compare} 
    \renewcommand{\arraystretch}{1.1}  
    \setlength\tabcolsep{6.0pt}  
    \resizebox{15.6cm}{!}{
        \begin{tabular}{cc|ccccc} 
            \toprule 
            \multicolumn{2}{c}{Models} & {\# Params} & {MAE $\downarrow$} & {Sim $\uparrow$} & {Rela $\uparrow$} & {Corr $\uparrow$}\\
            \midrule
            \multirow{2}{*}{Transformer~\cite{vaswani2017attention}} & {Original} & \multirow{2}{*}{9.55M} & {$1.454\times10^{-2}$} & {92.60\%} & {98.25\%} & {99.21\%} \\
            & {+ IA} & {} & {$6.937\times10^{-3}$} & {93.82\%} & {98.72\%} & {99.42\%} \\
            
            \cline{1-7}
            \multirow{2}{*}{LSTM~\cite{hochreiter1997long}} & {Original} & \multirow{2}{*}{7.11M} & {$2.197\times10^{-2}$} & {90.83\%} & {97.11\%} & {98.27\%} \\
            & {+ IA} & {} & {$3.827\times10^{-3}$} & {94.39\%} & {99.04\%} & {99.10\%} \\
    
            \cline{1-7}
            \multirow{2}{*}{Flowformer~\cite{wu2022flowformer}} & {Original} & \multirow{2}{*}{9.55M} & {$1.985\times10^{-2}$} & {93.09\%} & {97.91\%} & {99.46\%} \\
            & {+ IA} & {} & {$4.646\times10^{-3}$} & {94.54\%} & {98.98\%} & {99.53\%} \\
    
            \cline{1-7}
            \multirow{2}{*}{Non-stationary~\cite{liu2022non}} & {Original} & \multirow{2}{*}{7.13M} & {$8.577\times10^{-3}$} & {92.08\%} & {98.35\%} & {98.73\%} \\
            & {+ IA} & {} & {$4.298\times10^{-3}$} & {95.62\%} & {99.16\%} & {99.53\%} \\
    
            \cline{1-7}
            \multirow{2}{*}{\textbf{PaMMA-Net}} & {Original} & \multirow{2}{*}{\textbf{5.69M}} & {$6.711\times10^{-3}$} & {93.68\%} & {98.68\%} & {99.07\%} \\
            & {\textbf{+ IA}} & {} & {$\mathbf{2.661\times10^{-3}}$} & {\textbf{95.86\%}} & {\textbf{99.34\%}} & {\textbf{99.66\%}} \\
            \bottomrule
        \end{tabular}
    }
\vspace{-4mm}
\end{table*}

\subsection{Ablation study}
We conducted ablation experiments to validate the effectiveness of our proposed method. As shown in Table \ref{tab:ablation}, we first tested the model's performance using the complete method. Subsequently, we sequentially removed data augmentation, incremental design, positional embedding, and normalization to observe their individual contributions. Notably, data augmentation led to comprehensive performance improvements, particularly a 10\% relative reduction in MAE. The incremental design significantly enhanced performance on a higher baseline, with an approximate 1.5\% increase in Sim and a 49\% decrease in MAE. Positional embedding, a widely adopted technique, also demonstrated performance gains in the regression task of discharge prediction. Given the substantial differences in dynamic ranges among various signal channels, normalization played a crucial role in discharge prediction. In fact, the model without normalization failed to converge properly, resulting in poor prediction performance. This underscores the importance of normalization in our context.

\begin{table}[htbp]
    \centering
    \vspace{1mm}
    \caption{Results of the ablation on PaMMA-Net. The effectiveness of several components is validated by sequentially removing these components and testing the model's performance.}
    \label{tab:ablation} 
    \renewcommand{\arraystretch}{1.1}
    \setlength\tabcolsep{12.0pt}
    \resizebox{13cm}{!}{
        \begin{tabular}{c|cccc}
            \toprule 
            {PaMMA-Net} & {MAE $\downarrow$} & {Sim $\uparrow$} & {Rela $\uparrow$} & {Corr $\uparrow$} \\
            \midrule
            {Baseline} & {$3.074\times10^{-3}$} & {95.45\%} & {99.19\%} & {99.43\%} \\
            {-Data augmentation} & {$3.417\times10^{-3}$} & {95.14\%} & {99.13\%} & {99.32\%} \\
            {-Incremental design} & {$6.711\times10^{-3}$} & {93.68\%} & {98.68\%} & {99.07\%} \\
            {-Position embedding} & {$1.069\times10^{-2}$} & {93.13\%} & {98.11\%} & {99.14\%} \\
            {-Normalization} & {$8.044\times10^{-2}$} & {87.43\%} & {97.19\%} & {78.78\%} \\
            \bottomrule
        \end{tabular}
    }
\vspace{-1mm}
\end{table}
 
\subsection{Case study}
\begin{figure}[t!]
    \centering
    \includegraphics[width=15.6cm]{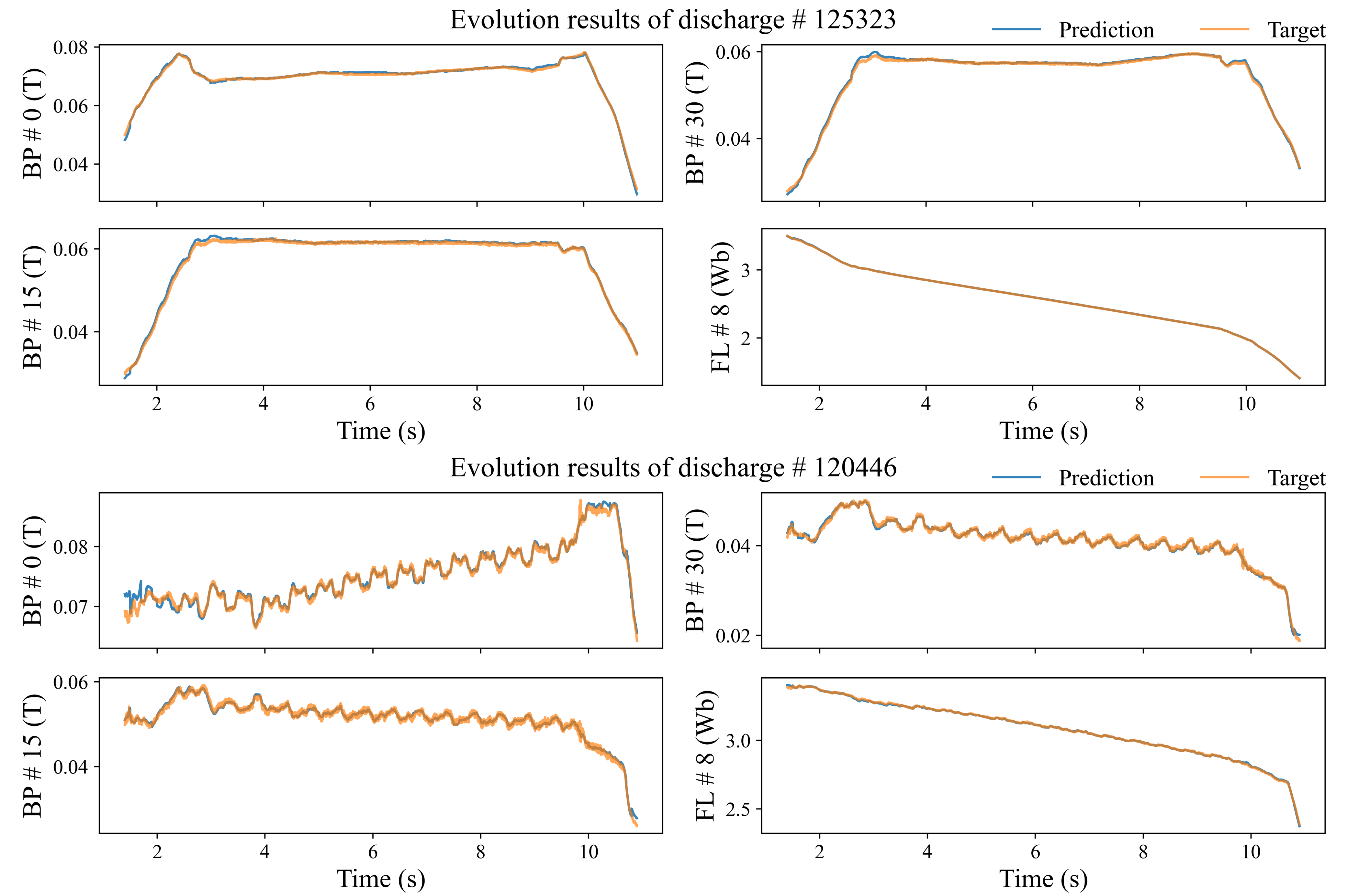}
    \caption{Examples of the non-autoregressive evolution outcomes throughout the entire discharging process.}
    \vspace{-3mm}
    \label{fig:discharge}
\end{figure}

In this section, the predictive capability of the model is intuitively demonstrated by showcasing its predictions on an entire shot. Although the maximum length for a single inference of the proposed model is limited to 1000 time steps, the entire shot prediction could be obtained by employing a sliding window on real experimental data and concatenating the prediction results (hereinafter referred to as the non-autoregressive prediction method). Fig. \ref{fig:discharge} presents an illustrative example of this process.

Fig. \ref{fig:discharge} presents examples of magnetic measurement evolution using the PaMMA-Net for discharge \# 125323 and \# 120446. Each case comprises four subplots, where three depict BP for different channels, and one showcases FL for a specific channel. In each subplot, the horizontal axis represents time, while the vertical axis denotes the measured values. The yellow line represents the experimental data, and the blue line indicates the model's predictions. Firstly, a notable advantage of the proposed surrogate model is its capability to predict the entire shot, encompassing both the rising and falling phases. In contrast, some response models are limited to predictions within the flat-top phase, thereby highlighting that the proposed model offers a more comprehensive predictive insight. Secondly, beyond matching the real values in overall trends, the model also exhibits proficiency in predicting abrupt changes during the flat-top phase. This is primarily attributed to the model's establishment of a relationship between real-time inputs and magnetic measurements, effectively reflecting mutations in real-time inputs into the magnetic measurements.

To showcase the details, only representative channels from the total 72 channel outputs are plotted here. The full-channel prediction results of the model across more shots are presented in the appendix.

\begin{figure}[t!]
    \centering
    \includegraphics[width=12cm]{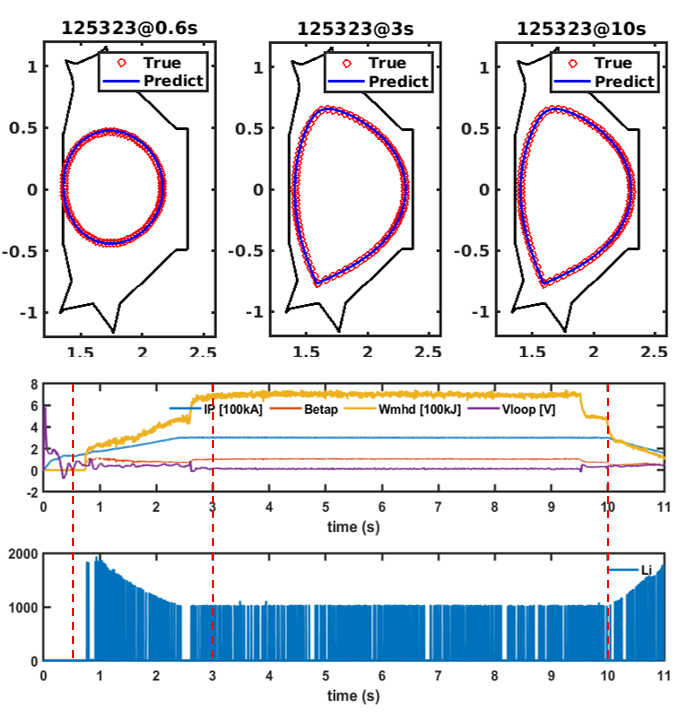}
    \caption{The evolution of LCFS derived from the combination of magnetic measurement evolution and equilibrium reconstruction.}
    \vspace{-3mm}
    \label{fig:LCFS}
\end{figure}

To further validate the evolutionary accuracy of PaMMA-Net, we also conducted equilibrium reconstruction on the evolution results to obtain the predicted plasma boundary. A comparison between the predicted boundary and the actual boundary is presented in Fig. \ref{fig:LCFS}. It is evident that the predicted boundary aligns well with the actual boundary, demonstrating that the model's magnetic measurement evolution results can be generalized to a broader range of plasma discharge parameters.

\subsection{Auto-regressive evolution}
\begin{figure}[t!]
    \centering
    \includegraphics[width=15.5cm]{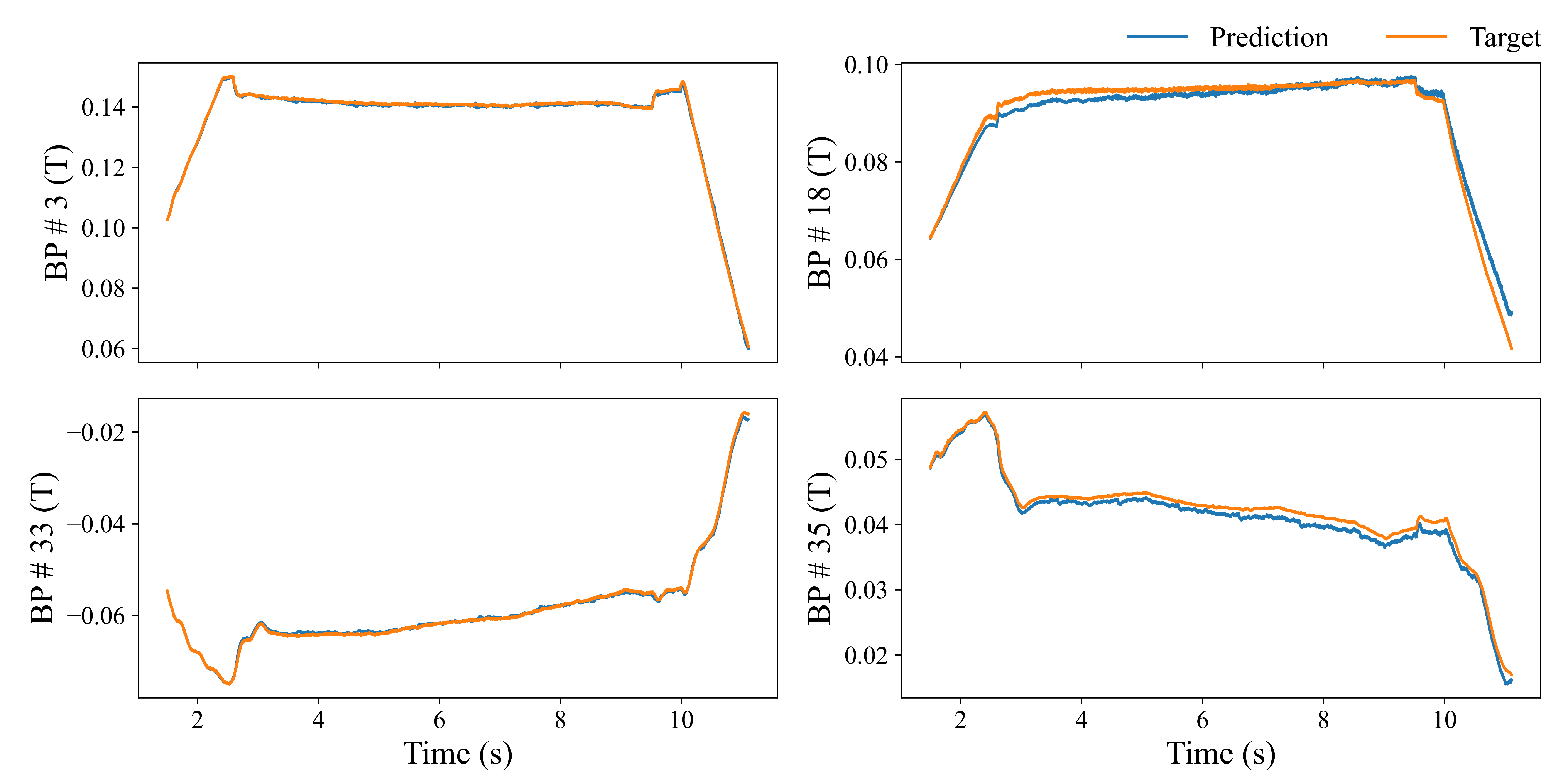}
    \caption{An example of the auto-regressive evolution outcomes throughout the entire discharging process.}
    \vspace{-3mm}
    \label{fig:autoreg}
\end{figure}

In this section, we explore the model's capability to predict the entire shot using an autoregressive approach. This means that the model can predict all subsequent magnetic measurements solely based on the initial 1000ms of probe signals during the rising phase. Of course, the known inputs, which include macroscopic variables and control variables, is provided in real-time. This is no different from the input required for traditional equilibrium evolution. An example of the model's autoregressive prediction result for the entire shot is shown in Fig. \ref{fig:autoreg}, which displays the prediction results for four channels.

Although our model has not undergone additional optimization for autoregressive prediction, we have found that its performance in predicting the entire shot using an autoregressive approach is acceptable. While the results are somewhat inferior to those obtained through non-autoregressive prediction, especially beyond 2000ms, the prediction errors do not diverge but remain within a certain range. This indicates that the model has reasonably modeled the relationship between real-time inputs and measured outputs. It demonstrates a strong robustness against disturbances in predicting magnetic measurements.

\section{Conclusion}
In this paper, PaMMA-Net is proposed for tokamak magnetic measurement evolution. By integrating knowledge from plasma physics and deep learning, innovative incremental design and data augmentation methods are introduced to effectively enhance the performance of evolution. When compared with various classical network architectures, PaMMA-Net achieves the best evolutionary results. The model's single-step inference capability implies its potential for offline interaction with controllers.

\section*{References}

\bibliographystyle{iopart-num}
\bibliography{cite}

\end{document}